\documentclass [11pt, letterpaper]{article}
\pdfoutput=1 
\usepackage{fullpage}
\usepackage{amsmath}
\usepackage{amssymb}
\usepackage{graphicx}
\usepackage{mathrsfs}
\usepackage{wrapfig}
\usepackage{boxedminipage}
\usepackage{epsfig}
\usepackage{setspace}
\usepackage{subfigure}
\usepackage{float}
\usepackage{hyperref}
\usepackage{enumerate}
\usepackage{cite}
\usepackage[font=footnotesize,labelfont=rm]{caption}

\begin{document}

\begin{flushright}
{\tt arXiv:1405.5941}
\end{flushright}

{\flushleft\vskip-1.35cm\vbox{\includegraphics[width=1.25in]{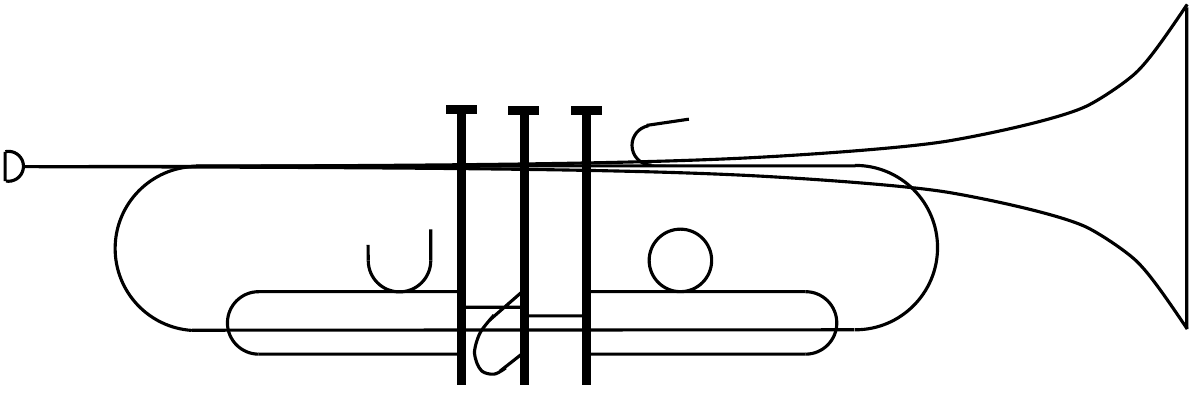}}}

\bigskip
\bigskip
\bigskip
\bigskip

\bigskip
\bigskip
\bigskip 
\begin{center} 

{\Large\bf Thermodynamic Volumes }

\bigskip

{\large\bf for  }
\bigskip

{\Large\bf AdS--Taub--NUT and AdS--Taub--Bolt\footnote{Original Title: ``Quantum Gravity and the Nuts and Bolts of Thermodynamic Volumes''.  To avoid confusion, the  title is adjusted here to match the title of the published version in the journal {\it Classical and Quantum Gravity}.}}

\end{center}

\bigskip\bigskip  \bigskip \bigskip \bigskip

\centerline{\bf Clifford V. Johnson}

\bigskip
\bigskip
\bigskip

  \centerline{\it Department of Physics and Astronomy }
\centerline{\it University of
Southern California}
\centerline{\it Los Angeles, CA 90089-0484, U.S.A.}

\bigskip

\centerline{\small \tt johnson1,  [at] usc.edu}

\bigskip
\bigskip


\begin{abstract} 
\noindent In theories of semi--classical quantum gravity where the cosmological constant is considered a thermodynamic variable, the gravitational mass of a black hole has been shown to correspond to the enthalpy of the thermodynamic system, rather than the energy. We propose that this should be extended to all spacetime solutions, and consider the meaning of  this extension of gravitational thermodynamics  for the Taub--NUT and Taub--Bolt geometries in four dimensional locally anti--de Sitter spacetime. We present formulae for   their thermodynamic volumes. Surprisingly, Taub--NUT has negative volume, for which there is a natural dynamical explanation in terms of the process of formation of the spacetime. A special case corresponds to pure AdS$_4$ with an $S^3$ slicing. The same dynamical setting can explain the negative entropy known to exist for these solutions for a range of parameters.  
\end{abstract}
\newpage \baselineskip=18pt \setcounter{footnote}{0}

\section{Extended Gravitational  Thermodynamics}
%

Black hole thermodynamics\cite{Bekenstein:1973ur,Bekenstein:1974ax,Hawking:1974sw,Hawking:1976de}, which relates the mass $M$, surface gravity $\kappa$, and area $A$ of a black hole to the energy $U$, temperature $T$, and entropy $S$, according to:
\begin{equation}
\frac{M}{G}=U\ ,\quad T=\frac{\kappa}{2\pi}\ , \quad S=\frac{A}{4G}\ ,
\end{equation}
($G$ is Newton's constant) has recently been extended\footnote{For a selection of references, see refs.\cite{Caldarelli:1999xj,Wang:2006eb,Sekiwa:2006qj,LarranagaRubio:2007ut,Kastor:2009wy,Dolan:2010ha,Cvetic:2010jb,Dolan:2011jm,Dolan:2011xt}, including the reviews in refs.\cite{Dolan:2012jh,Altamirano:2014tva}. See also the early work in refs.\cite{Henneaux:1984ji,Teitelboim:1985dp,Henneaux:1989zc}.} to include black hole counterparts for the pressure $p$ and volume $V$. The  cosmological constant of the spacetime in question supplies the pressure through the relation $p=-\Lambda/8\pi G$, while the thermodynamic volume $V$ is a derived quantity that in static cases is associated with the volume occupied by the black hole itself. (We will use geometrical units where $c,\hbar,k_{\rm B}$ have been set to unity.)  The formalism works in multiple dimensions, and our remarks will apply to those situations too, although  we will  study  four--dimensional  examples in this paper. The black holes may have other parameters such as gauge charges $q_i$ and angular momenta $J_i$, and these, with their conjugates the potentials  $\Phi_i$ and angular velocities $\Omega_j$, enter additively into the First Law in the usual manner. 

In the presence of a variable pressure $p$, the extension shifts\cite{Kastor:2009wy} the identification of the mass $M$  from determining the internal energy $U$ to  setting the {\it enthalpy}: $M/G=H\equiv U+pV$. So the First Law now becomes:
\begin{equation}
dM=TdS+Vdp+ \Phi dq+ \Omega dJ\ ,
\end{equation}
in four dimensions with an electric charge and rotation. When $p$ is removed from the list of variables, we return to the usual situation.

In the case of static black holes, the thermodynamic volume $V$ is simply the ``geometric'' volume constructed by naive use of the radius of the black hole horizon\footnote{This result agrees with the definition of the volume of a static black hole proposed in ref.\cite{Parikh:2005qs}.}. For example, in four dimensions,  for a Schwarzschild black hole with horizon radius  $r_h$, we have 
\begin{equation}
V=V_{\rm sch}=\frac{4}{3}\pi r_h^3\ .
\label{eq:naive}
\end{equation}
In general, the thermodynamic entropy is not equal to the naive geometric volume\cite{Cvetic:2010jb}.

Enthalpy is very natural here\cite{Kastor:2009wy}: The cosmological constant is a spacetime energy density of  $-p=\Lambda/8\pi G$ per unit volume. Forming a black hole of volume $V$ requires cutting out a region of spacetime of that volume, at cost $pV$,  and this energy of formation is naturally captured by the enthalpy. 

It seems an important fact  that once we have dynamical $p$, and its conjugate  $V$, we ought to  explore the full range of thermodynamic physics that having such variables available affords us. For example, as pointed out in ref.\cite{Johnson:2014yja}, it is natural to consider building heat engines in this gravity setting, where mechanical work {\it i.e.} actual changing of thermodynamic volumes, can be performed\footnote{That paper also discussed  the consequences of the extended thermodynamics, and the heat engine proposal, for dual field theories {\it via} holography\cite{Maldacena:1997re,Gubser:1998bc,Witten:1998qj,Witten:1998zw} in the case of negative cosmological constant.}.  A  key idea to take away here is that one should be able  to consider a particular spacetime solution to be the result of a (gravitational) thermodynamic process, and since enthalpy  seems to be a natural and central quantity, we should trace not just energy $U$ but pay attention to the processes that give rise to a particular energy of formation. We will take this idea seriously in this paper. 

One of the main points of this paper is to emphasise  that if we are to take the extended thermodynamics seriously, it should apply not just to black hole spacetimes, but {\it all} spacetimes, and the consequences should be explored. Indeed, for negative cosmological constant at least, holography\cite{Maldacena:1997re,Gubser:1998bc,Witten:1998qj,Witten:1998zw} would appear to demand this. (See ref.\cite{Johnson:2014yja} for a discussion of how this extended thermodynamics fits with holography.) In particular, there are  ways of assigning thermodynamic quantities such as temperature and entropy to spacetimes that do not have horizons~\cite{Hawking:1998jf}, using the same semi--classical quantum gravity calculus that endows black holes with their thermodynamic properties. It seems therefore neglectful (at best) to not consider what these spacetimes' thermodynamic properties are in this extended scheme, being careful to interpret these properties in the light of volume being dynamical.  

Two examples that spring to mind are the Taub--NUT \cite{Taub:1950ez,Newman:1963yy} and Taub--Bolt \cite{Page:1979aj} spacetimes, which have a metric of the following form for negative cosmological constant\cite{Page:1985bq,Page:1985hg}:
\begin{equation}
\label{eq:metric}
ds^2=F(r)(d\tau+2n\cos\theta d\phi)^2+\frac{dr^2}{F(r)}+(r^2-n^2)(d\theta^2+\sin^2\!\theta d\phi^2)\ ,
\end{equation}
where
\begin{equation}
\label{eq:metric-function}
F(r)\equiv\frac{(r^2+n^2)-2mr+\ell^{-2}(r^4-6n^2r^2-3n^4)}{r^2-n^2}\ ,
\end{equation}
with the cosmological constant $\Lambda=-3/\ell^2$. Time $\tau$ here is Euclidean with period $\beta=8\pi n$ (in order to ensure the invisibility of Misner strings\cite{Misner:1963fr}), and the parameters $m$  and $n$, and the range of $r$  will have certain restrictions (depending upon whether we are Taub--NUT or Taub--Bolt) that we will review below. The spaces are asymptotically AdS$_4$, generically; The topological $S^3$ formed by the $\tau$ circle fibred over the $S^2$ of $\theta$ and $\phi$ is squashed for general $n$. We will study each  case in the next sections, with interesting results for the extended thermodynamics.

\section{AdS--Taub--NUT}
\label{sec:taub-nut}
\subsection{General $n$}
For the metric given in equations~(\ref{eq:metric}) and~(\ref{eq:metric-function}) the circle parameterized by  $\tau$ is Hopf fibred over the~$S^2$ with first Chern class $n$. The  vector $\partial_\tau$ that generates the $U(1)$ of translations along the circle will have  fixed points (where the fibre degenerates) at some $r$ when $F$ vanishes.  The difference between a nut and a bolt is whether the fixed point set is zero dimensional or  two dimensional\cite{Gibbons:1979xm}. The former case is the nut, and is at $r=r_n=n$ where the $S^2$ is of zero size. There, $F(r=r_n)=0$.  There is an additional condition on $F$ at this point  to ensure that there's no conical singularity: $F^\prime(r=r_n)=1/2n$. These conditions on the parameters in the metric were made explicit in ref.\cite{Chamblin:1998pz}, and for completeness we quote them here.  The mass parameter $m$ becomes 
\begin{equation}
\label{eq:taub-nut-mass}
m_n=n-\frac{4n^3}{\ell^2}\ ,
\end{equation}
simplifying the metric function $F(r)$,
and we have that $n\leq r \leq +\infty$. 
In fact the action  of this spacetime was computed in refs.\cite{Mann:1999pc,Emparan:1999pm} to be:
\begin{equation}
I=\frac{4\pi n^2}{G}\left(1-\frac{2n^2}{\ell^2}\right)\ ,
\end{equation}
from which the mass (divided by $G$) given above emerges as $\partial_\beta I$, (recall that $\beta=8\pi n$) and the entropy as
\begin{equation}
\label{eq:taub-nut-entropy}
S_n=\beta\frac{\partial I}{\partial \beta} - I = \frac{4\pi n^2}{G}\left(1-\frac{6 n^2}{\ell^2}\right)\ .
\end{equation}
Consider the extended thermodynamics, where the cosmological constant defines a dynamical variable, the pressure  $p=-\Lambda/8\pi G=3/(8\pi\ell^2 G)$. As recalled in the previous section, we should interpret $m_n/G$ not as the energy $U$ but the  enthalpy $H=U+pV$ of the gravitational  thermodynamics, where $H(S,p)$ has $S$ and $p$ as its natural variables, and the First Law yields:
\begin{equation}
\label{eq:first-law}
dH=TdS+Vdp\ .
\end{equation}
 So  we have:
\begin{equation}
\label{eq:taub-nut-enthalpy}
H_n(S,p)=\frac{1}{G}\left(n-\frac{32\pi Gn^3}{3}p\right)\ , 
\end{equation}
and the full $(S_n,p)$  dependence of $H(S_n,p)$ is implied through the elimination of $n$ between this and equation~(\ref{eq:taub-nut-entropy}). As a check, we can recover the temperature through 
\begin{equation}
T=\left.\frac{\partial H_n}{\partial S}\right|_p =\left.\frac{1}{G}\frac{\partial n}{\partial S}\right|_p\left(1-32\pi G n^2 p\right)  = \frac{1}{8\pi n}\ ,
\end{equation}
since, from~(\ref{eq:taub-nut-entropy}):
\begin{equation}
\left.\frac{\partial S}{\partial n}\right|_p = \frac{8\pi n}{G}(1-32\pi G  n^2p)\ .
\end{equation}
Now we are ready for something new. We define the thermodynamic volume through the relation $V=\left.(\partial H/\partial p)\right|_S$, and get:
\begin{equation}
\label{eq:taub-nut-volume}
V_n=\left.\frac{\partial H_n}{\partial p}\right|_S= -\frac{32\pi n^3}{3}+\left.\frac{\partial n}{\partial p}\right|_S\left(\frac{1}{G}-32\pi n^2p\right)=-\frac{8\pi n^3}{3}\ ,
\end{equation}
since, from~(\ref{eq:taub-nut-entropy}):
\begin{equation}
\left.\frac{\partial n}{\partial p}\right|_S = {8\pi n^3}\left(\frac{1}{G}-32\pi   n^2p\right)^{-1}\ .
\end{equation}
Equation~(\ref{eq:taub-nut-volume}) is our first new result, and we should pause to understand it. First, we note that it is negative, which is certainly puzzling for a volume! We will interpret the sign shortly. Sign aside, we should also note that this is a new and rather extreme example of a case where  the thermodynamic volume and the naive geometric volume differ, joining the example of the rotating black hole\cite{Cvetic:2010jb}. Here, since the $S^2$ at the origin is of zero radius, the naive geometric volume entirely vanishes! 

Before proceeding, we should check that our result is consistent with scaling in four dimensions, by seeing if our thermodynamic quantities satisfy a Smarr relation\cite{Smarr:1972kt}. Recalling that the temperature is $T=1/(8\pi n)$, the expressions~(\ref{eq:taub-nut-enthalpy}),~(\ref{eq:taub-nut-entropy}) and~(\ref{eq:taub-nut-volume}) can be combined to show that:
\begin{equation}
\label{eq:smarr}
\frac{H}{2}-TS+pV=0\ ,
\end{equation}
which is satisfying. In fact, it was the cosmological constant modification of the Smarr relation that inspired the suggestion\cite{Kastor:2009wy} that the black hole mass should be treated as an enthalpy.  Note that we can also derive an expression for the internal energy of Taub--NUT in this picture by defining $U=H-pV$, giving:
\begin{equation}
\label{eq:taub-nut-energy}
U_n=\frac{n}{G}\left[1-\frac{3n^2}{\ell^2}\right]\ .
\end{equation}

To help interpret the negative sign of the  volume~(\ref{eq:taub-nut-volume}), recall the case of Schwarzschild that we reviewed in the previous section. There, the thermodynamic volume $V_{\rm Sch}$ recovered was the flat space volume of a ball whose radius was that of the horizon. Recall that the enthalpy is the energy of the system plus the energy needed to make the system,  so  one imagines starting with an empty spacetime with coordinate $r$ running from 0 to $\infty$, and  cosmological constant $\Lambda$. Making the black hole  involves cutting out the volume $V_{\rm Sch}$. Cutting out this chunk of spacetime to insert the black hole costs energy $-\Lambda V_{\rm Sch}/8\pi G=pdV  = pV_{\rm Sch}$.  

This is a core idea that we want to refine and exploit in what is to follow, so let us state clearly what we have in mind. In that static black hole case,   we should think of forming the whole black hole solution --- our ``system'' --- {\it via} a thermodynamic process. In this process, the system does work on the universe (pushing against it and reducing its volume) in the process of its creation,  and this is what went into the enthalpy. 

We should make exactly the same interpretation here for Taub--NUT. We imagine starting with empty space again, with a coordinate $r$ running from 0 to $\infty$ as before. We must now adjust the volume in order to form our Taub--NUT geometry and then repair it again where the spacetime ``starts'' at $r=n$. There is a subtle difference with the black hole case. There, the $S^2$ where the radial coordinate  begins at $r=r_h$ is at finite size, which is consistent with having cut something out of the spacetime. Here,  $r$ begins away from zero again, at $r=n$, but the $S^2$ is  at zero size too. This, we propose,  should be interpreted not as cutting out some volume, but instead {\it adding   some in}. In other words, in our dynamical language, the  {\it universe} does work to create the system (our Taub--NUT solution), not the other way around. So the contribution to the enthalpy is negative: $-p|V_n|$. So there is not really a negative volume for our system if we interpret appropriately in this dynamical setting. Notice that since there's no geometrical counterpart for this volume left anywhere in the spacetime the formalism  neatly avoids any puzzling issues with sign interpretations.  A special case of all this is the nut charge value $n=\ell/2$. This is nothing more than AdS$_4$ in an unusual slicing, and gives us a familiar example to understand all this with. We study it in  subsection~\ref{eq:ads4}.

\subsection{An Embedding Visualisation}
\begin{wrapfigure}{R}{0.5\textwidth}
{\centering
\includegraphics[width=2.8in]{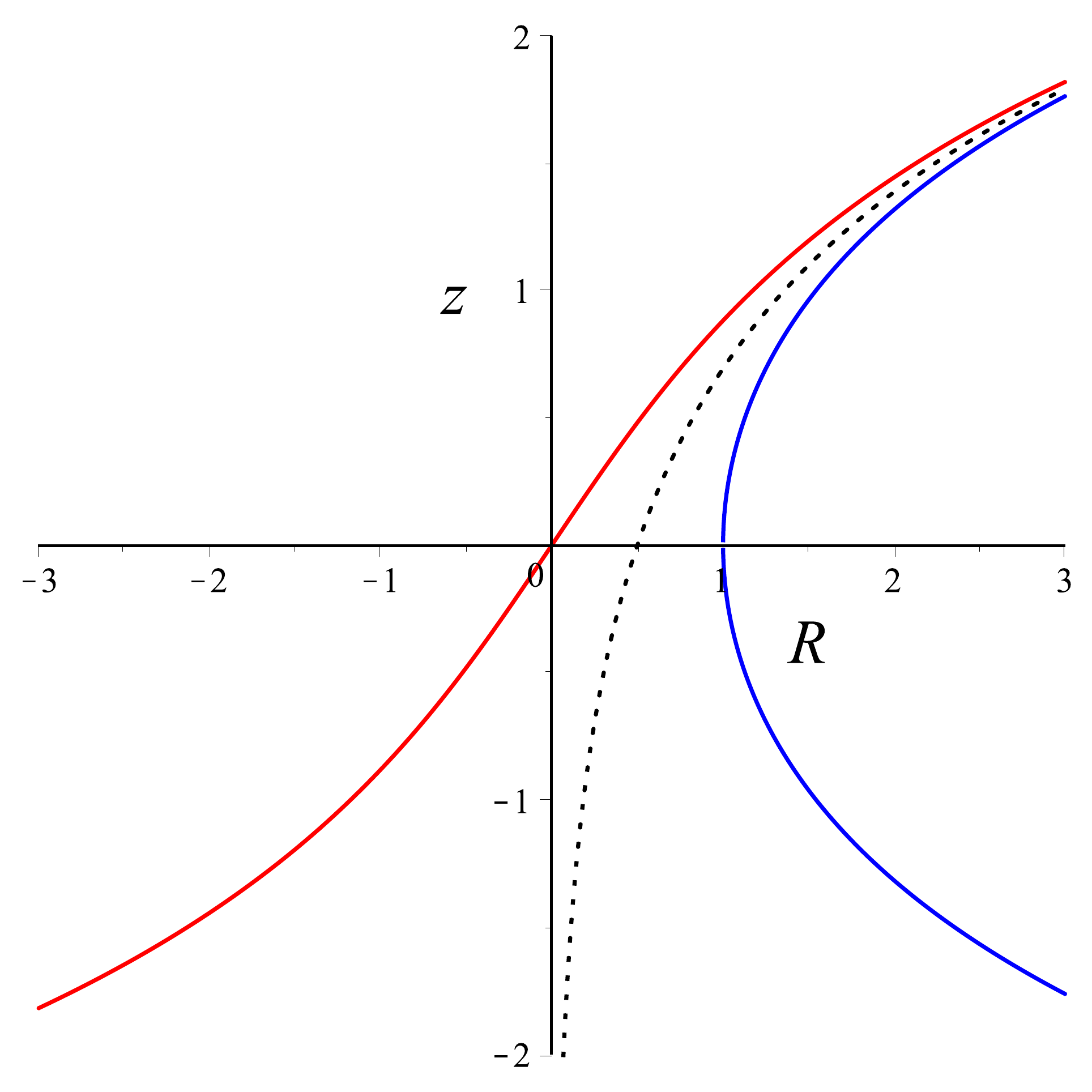} 
   \caption{\footnotesize   The magnitude of the function $z(R)$ giving the embedding of our model space~(\ref{eq:modelspace}) into a  space with radial coordinate $R$ and vertical coordinate $z$.  The trivial case $R=r$ is represented by the axis $z=0$.  The central dotted line is the asymptote $z(R)=n\ln(2R)$.   The  solid curve approaching it  from beneath  is the  $R^2=r^2+n^2$ case. The solid curve approaching from above is the $R^2=r^2-n^2$ case (in fact, $z(R)$ is purely imaginary in this case). For the  curves we used $n=1$. See  text for interpretation.  }  \label{fig:embedding}
 
}
\end{wrapfigure}

Let us pause to study  a possibly helpful way of visualising the claimed interpretation of the volume,  through an embedding of the spatial metric into a space of one  dimension higher. Our focus is the behaviour of the $S^2$s near the origin of coordinates. 
Let us compare three cases: (a) A factor $r^2-n^2$, (b) an $r^2$ factor and also (c) a factor $r^2+n^2$, the latter being  analogous to the black hole  case of finite area horizon at the origin of the radial spatial coordinate. A simple model metric is 
\begin{equation}
d{\tilde s}^2= dr^2+(r^2\pm n^2)d\Omega_2^2\ ,\quad{\rm where}\quad d\Omega_2^2=d\theta^2+\sin^2\!\theta d\phi^2\ ,
\label{eq:modelspace}
\end{equation}
where $n=0$ simply gives $\mathbb{R}^3$.
 We can  embed our model space into a higher dimensional 	``cylindrical'' four dimensional metric
$ds^2= dR^2+R^2d\Omega_2^2+dz^2,$
where $z=z(r)$ and $R=R(r)$. 
The embedding implies the following relations for the coordinates:
\begin{equation}
R^2=r^2\pm n^2\ ,\qquad \left(\frac{dz}{dr}\right)^2+\left(\frac{dR}{dr}\right)^2 = 1\ .
\end{equation}
Some algebra gives:
\begin{equation}
z(r)=({\pm1})^\frac12n\ln(r+\sqrt{r^2\pm n^2})\ .
\end{equation}
Solving for $z(R)$, in  case (c), we get $z(R)=n\ln(R\pm\sqrt{R^2-n^2})$ and in  case (a) we have $z(R)=in\ln(R+\sqrt{R^2+n^2})$. The latter is imaginary\footnote{We thank Wolfgang Mueck for pointing out an error in an earlier version of this discussion, and  an anonymous referee for pointing out the signature change.}, telling us that instead we should embed this case into the space with Lorentzian signature obtained by continuing $z\to iz$. For the qualitative comparison we wish to make here, it is the overall dependence on $R$ that we are interested in comparing and contrasting, so we  plot the magnitude of $z(R)$ for the three cases together  in figure~\ref{fig:embedding}.
 
 The  case~(b), $n=0$, is just the horizontal axis $z=0$, the trivial embedding. The  cases (a) and (c) asymptotically agree at large~$R$, approaching   the asymptote $z=n\ln(2R)$ (the dotted line).  The  solid curve approaching from below  is  case (c) where the $S^2$s have factor $(r^2+n^2)$ and the solid curve approaching from above is the imaginary part of case (a) where the $S^2$s have factor $(r^2-n^2)$. 

In the  $n=0$ case  the space can go all the way down to $R=0$, where the $S^2$s shrink to zero size, which is natural for $\mathbb{R}^3$.  For the  lower curve, $R$ decreases to a minimum value,~$n$, and then increases positively again by going from one branch of the square root to the other\footnote{This corresponds to letting $r$  go negative,  and can be considered as constructing a wormhole solution, but here we will identify $r$ and $-r$.}. The $S^2$s never shrink to zero size.  By contrast, the top curve has $R$ decrease to zero again, allowing the $S^2$s to pinch down to zero size\footnote{Although for this case  the embedding function has finite derivative at the origin, the full nut solution would give a smooth space since a whole $S^3$ is shrinking and so the origin looks like that of ${\mathbb R}^4$. This also looks odd here because we are embedding into a cylindrical coordinate system which naturally focuses on $S^2$s.}.

We can see by comparing the three embeddings on the same plot  that they fill the embedding space rather differently, the case with $r^2+n^2$ filling less than the $r^2$ case, and the $r^2-n^2$ case filling more than the other two cases. It is in this (admittedly partly qualitative) sense that the Taub--NUT example corresponds to  an {\it increase} in volume of the environment (the universe, played here by the $r^2$ case) for its formation, giving a negative volume in the enthalpy, as we saw.

\subsection{A Naive Derivation of the Thermodynamic Volume}
\label{eq:naive-volume}
We can change tactics here and try to derive  our formula for the thermodynamic volume by doing  a naive computation designed to mimic what was done for static black holes (reviewed in the introduction).  The volume in equation~(\ref{eq:naive}) can   be obtained by computing the four--dimensional volume element from the (AdS) Schwarzschild metric, and simply taking only the spatial part $dV=r^2\sin\theta dr d\theta d\phi$. The volume is then obtained by  integrating from $r=0$ to $r=r_h$. In a sense,  $ds^2=dr^2+r^2d\Omega_2^2$ is the ``effective'' spatial metric yielding that volume.

Turning to our case, the same procedure on equations~(\ref{eq:metric}) and~(\ref{eq:metric-function}) gives spatial volume element $dV=(r^2-n^2)\sin\theta dr d\theta d\phi$. So the  effective spatial metric in our case is now:
$ds^2=dr^2+(r^2-n^2)d\Omega_2^2$, the case $(a)$ of the model metric of the previous subsection.  Now although the spacetime really begins at $r=n$,   by analogy with the previous paragraph we should integrate from $r=0$ to $r=n$,  giving:
\begin{equation}
V=4\pi\int_0^n(r^2-n^2)dr = \Biggl.\frac{4\pi}{3}(r^3-3n^2 r)\Biggr|_{r=0}^{r=n}=-\frac{8\pi n^3}{3}\ .
\label{eq:naive-integral}
\end{equation}
This is the $V_n$ we obtained earlier, in equation~(\ref{eq:taub-nut-volume})! This lends support to our idea expressed above that the ``added" volume for Taub--NUT is negative. This is not a proof, since this procedure for deriving the thermodynamic volume is heuristic at best. While the method works for both cases, it is  worth noting that  a key difference between (Euclidean)  Schwarzschild and  Taub--NUT  is that there is no distinguished  geometrical surface in Taub--NUT  to which we can associate a finite volume. Its origin  at $r=n$ is topologically~$\mathbb{R}^4$ instead of $\mathbb{R}^2\times S^2$.  The thermodynamic volume $V_n$ is highly non--geometric in this sense, since the naive geometric volume is zero, the volume of the~$S^2$ of vanishing size at the origin.

As we shall see, this naive procedure will reproduce the result  for the thermodynamic volume~$V_b$ that we will get  for the Taub--Bolt case  studied in section~\ref{sec:taub-bolt}.  In that case, as we will review,  the spacetime begins at some ``bolt radius'' $r_b>n$. So our naive procedure here suggests that to work out the volume we should do the integral in equation~(\ref{eq:naive-integral}) again, but now integrate~$r$ from zero  to $r_b$. The resulting expression is exactly what we will get by using the more careful  enthalpy procedure.

\subsection{$n=\ell/2$ and  AdS$_4$}
\label{eq:ads4}
For the case $n=\ell/2$, the AdS--Taub--NUT solution of the previous section is in fact AdS$_4$ with a non--trivial slicing. (The trivial case of AdS$_4$ here, having   $S^1\times S^2$ slices, arises from setting $n=0$.) Writing (as  in ref.\cite{Chamblin:1998pz}) $r=\rho+n$, the radial slicing is such that   the angle $\psi= \tau/\ell$ (period $4\pi$) and the $S^2$ angles $(\theta,\phi)$ form a round $S^3$, and the metric becomes:
\begin{equation}
ds^2 = \frac{{d\rho^2}}{f(\rho)} + f(\rho)\ell^2[(d\psi+\cos\theta d\phi)^2+ d\theta^2+\sin^2\!\theta d\phi^2]\ ,
\end{equation}
where
\begin{equation}
f(\rho)= \frac{\rho^2}{\ell^2}\left(1+\frac{\ell}{\rho}\right)\ ,
\end{equation}
which gives, once we put the boundary at finite distance ($y=\ell$) using the transformation $y^2/\ell^2=\rho/(\rho+\ell)$, the standard $S^3$ radial slicing of AdS$_4$ known as the Poincar\'e ball.

For this case,  $m_n=0$, and  therefore the enthalpy $H$ vanishes too. From the perspective of the extended thermodynamics we  have a new understanding of this. The enthalpy vanishes because there is a cancellation between the energy of the spacetime, $U_{\ell/2}$,  and the energy of formation, $-p|V_{n=\ell/2}|$, where:
\begin{equation}
|V_{n=\ell/2}| = \frac{\pi \ell^3}{3}\ , \quad U_{\ell/2} = \frac{\ell}{8G} \ .
\end{equation}
According to our dynamical interpretation of the previous section, $\pi\ell^3/3$ is the amount by which the volume of the universe increased in order to form the spacetime. 

It has already been noticed\cite{Emparan:1999pm,Mann:1999pc} that the non--vanishing entropy of AdS$_4$ for this time slicing  is negative: $S=-\pi\ell^2/2G$. This somewhat puzzling result can be given a physical interpretation in the context of the ideas presented here. There is no reason to suppose that the  process of creating the thermodynamic volume need be adiabatic. The negative entropy of the spacetime would result from some heat flow of magnitude $TS=\ell/8G$ out of the volume. In fact, putting in the magnitude of our volume, $\pi\ell^3/3$, we find that the heat flow precisely matches the work done: $pV=\ell/8G$, and so the internal energy was held constant in the formation process! A similar interpretation can be given to all the negative entropy cases following from equation~(\ref{eq:taub-nut-entropy}), {\it i.e.}, for $n>\ell/\sqrt{6}$.


\section{AdS--Taub--Bolt}
\label{sec:taub-bolt}
There is another way for the vector $\partial_\tau$ to degenerate, as already mentioned. It can have a two dimensional fixed point set, known as a bolt\cite{Page:1979aj,Gibbons:1979xm}.  So the metric function $F(r)$ vanishes at some radius~$r_b$  that is greater than $n$. Ensuring a smooth metric again, yields for the mass\cite{Chamblin:1998pz}:
\begin{equation}
\label{eq:bolt-enthalpy}
m_b=\frac{r_b^2+n^2}{2r_b}+\frac{1}{2\ell^2}\left(r_b^3-6n^2 r_b - 3\frac{n^4}{r_b}\right)\ ,
\end{equation}
with $r_b$ solving a quadratic equation, which gives $r_b$  in two branches:
\begin{equation}
\label{eq:arbyquadratic}
6nr_b^2-\ell^2 r_b -6n^3+2n\ell^2=0\ ;\quad  r_{b\pm}=\frac{\ell^2}{12 n }\left(1\pm\sqrt{1-48\frac{n^2}{\ell^2}+144\frac{n^4}{\ell^4}}\right)\ , 
\end{equation}
with $n$ restricted to be no greater than $n_{\rm max}$ to have real $r_b$ greater than $n$:
\begin{wrapfigure}{R}{0.4\textwidth}
{\centering
\includegraphics[width=2.6in]{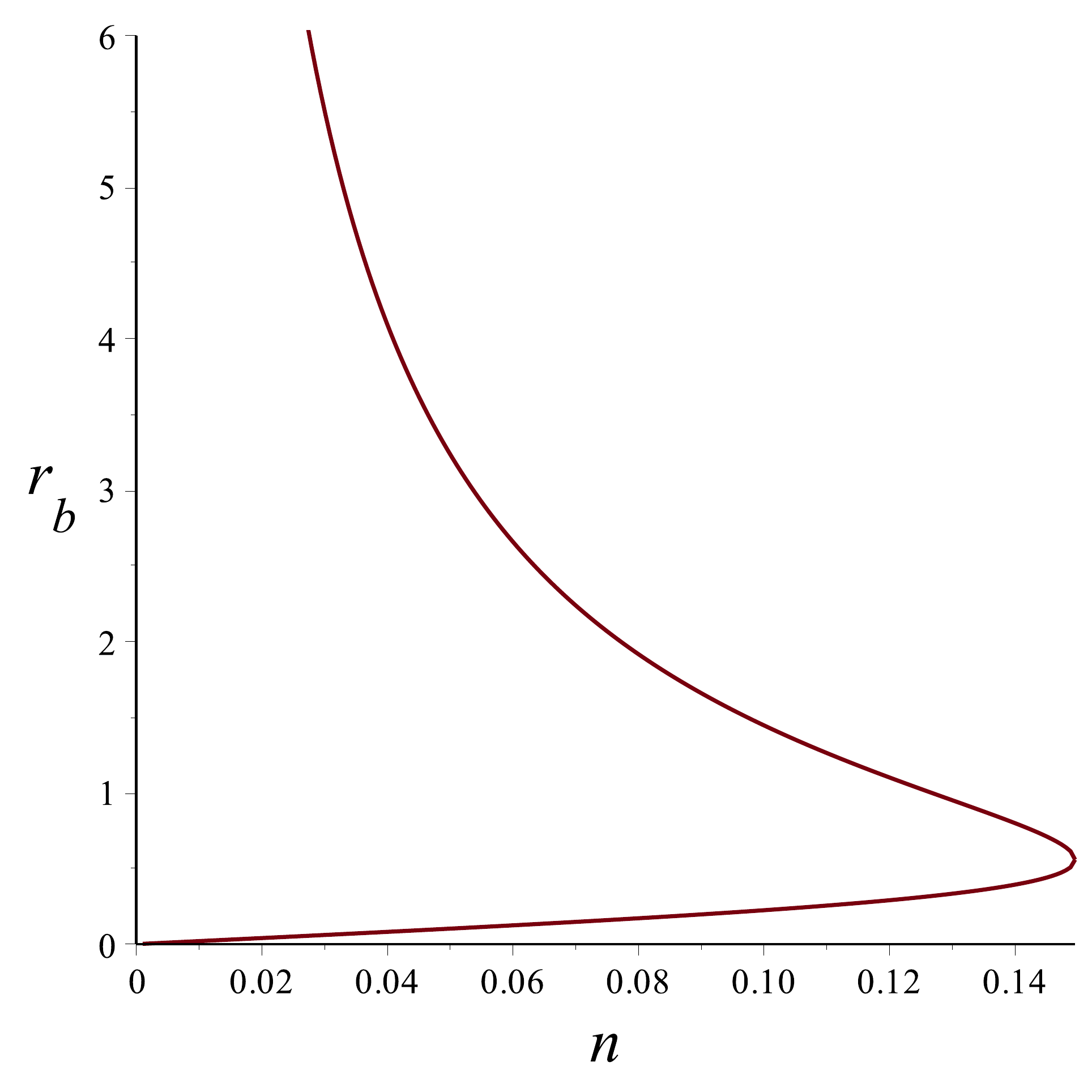} 
   \caption{\footnotesize   The available bolt radii, the merger of an upper and a lower branch that connect at $n_{\rm max}\simeq 0.149\ell$. Here we chose $\ell=1$.}  \label{fig:arbyplot}
}
\end{wrapfigure}
\begin{equation}
n_{\rm max}=\left(\frac16-\frac{\sqrt{3}}{12}\right)^{\frac12}\ell\ .
\end{equation}

The function $r_b(n)$ is  plotted in figure~\ref{fig:arbyplot}.  In fact, the mass also follows from the form of the action derived in ref.\cite{Emparan:1999pm}:
\begin{equation}
I=\frac{4\pi n }{G\ell^2}\left(\ell^2m+3n^2r_+-r_+^3\right)\ ,
\end{equation}
where for this case we substitute $m=m_b$ and $r_+=r_b$. (The Taub--NUT action of the previous section comes from substituting $m=m_n$ and $r_+=n$ into this expression.) We can go further and derive the entropy through the usual thermodynamic relation $S=(\beta\partial_\beta-1)I$, giving:
\begin{eqnarray}
\label{eq:taub-bolt-entropy}
 S_b&=&\frac{4\pi n}{G}\left(m_b-\frac{3n^2 r_b}{\ell^2}+\frac{r_b^3}{\ell^2}\right)\\
 &=& \frac{4\pi n}{G}\left[\frac{r_b^2+n^2}{2r_b}+\frac{4\pi Gp}{3}\left(3r_b^3 -12 n^2 r_b-\frac{3n^4}{r_b}\right)\right]\ .  \nonumber
\end{eqnarray} 
Let us next turn to the issue of the enthalpy and the thermodynamic volume for this spacetime in the extended thermodynamics. We propose that the bolt mass  defines an enthalpy $H_b=m_b/G$ in the extended thermodynamics:
\begin{equation}
 H_b(S_b,p)= \frac{1}{G} \frac{r_b^2+n^2}{2r_b}+\frac{4\pi}{3}p\left(r_b^3-6n^2 r_b - 3\frac{n^4}{r_b}\right)\ ,
\end{equation} 
where the explicit ($S_b,p$) dependence follows from the fact that $m_b$  has an $r_b$, $n$, and $\ell$ dependence, as does $S_b$. Given that $r_b$ also depends on $n$ and $\ell$, there are enough relations given the number of variables to ensure a well--defined $H(S,p)$, as in the Taub--NUT case. 

Note that it is automatic that we can recover the temperature $T=1/\beta$ through the relation $T=\left.\partial H/\partial S\right|_p$ given that the derivatives used to recover $m_b/G$ and $S_b$ from the action were already performed by treating~$\ell$ (and hence $p$) as a constant. Taking an additional $\beta$ derivative on $H_b$ and on $S_b$ and using the chain rule then completes the demonstration. 

Now we are ready to carry out the procedure done in the Taub--NUT case, holding $S$ fixed so that we can define the volume {\it via}: $V=\left.(\partial H/\partial p)\right|_S$. We need the $p$ derivatives of $n$ and $r_b$ for fixed $S_p$, and some tenacity in our algebra. This is where the Smarr relation~(\ref{eq:smarr}), which follows entirely from scaling and so must be true here too, can be used to simplify matters considerably. Since we have especially simple expressions for $T$ and $p$, and we have $H_b$ and $S_b$ worked out already, we can extract the volume as $V_b=(TS_b-H_b/2)/p$, and a little algebra yields:
\begin{equation}
\label{eq:taub-bolt-volume}
V_b = \frac{4\pi}{3}\left(r_b^3-3n^2 r_b\right)\ ,
\end{equation}
a pleasingly simple formula. Notice that it matches the formula that we anticipated at the end of subsection~\ref{eq:naive-volume} by using an heuristic  argument!

Some checks are in order.  First, notice that our last few manipulations did not  use the explicit form of the Taub--Bolt equation~(\ref{eq:arbyquadratic}) for $r_b$, and so this result is in fact true for the Taub--NUT case too, as can be seen by setting the radius to $n$, upon which we recover equation~(\ref{eq:taub-nut-volume}). (This also fits with with our observation from subsection~\ref{eq:naive-volume} that $V_b$ and $V_n$ can both come from the same  integral formula.)
Second, notice that this result is manifestly positive, since the closest~$r_b$ can get to $n$ is asymptotically, at small $n$: $r_b=2n+18n^3/\ell^2+O(n^5)$ (the lower branch solutions in figure~\ref{fig:arbyplot}). Third, for large $r_b$ (the upper branch solutions in figure~\ref{fig:arbyplot}, also at small~$n$), the volume becomes the naive geometrical volume  $V_b\sim 4\pi r_b^3/3$. Away from the large $r_b$ limit, we see that we have found another example of a thermodynamic volume that is different from the naive geometric volume. This time instead of being due to rotation as in ref.\cite{Cvetic:2010jb}, the discrepancy is due to the presence of nut charge.
 
 The volume $V_b$ has $\ell$ dependence through that of $r_b$, while the volume for Taub--NUT, $V_n$ does not. It is amusing to note that in the asymptotically (locally) flat case,  the limit  $\ell\to\infty$ ({\it i.e.,}  $\Lambda=0$), the thermodynamic volumes for Taub--NUT and Taub--Bolt become equal in magnitude (although opposite in sign in the sense discussed earlier), because $r_b\to 2n$ and so: 
 \begin{equation}
\label{eq:special-volume-limit}
\biggl.V_b\to \frac{8\pi n^3}{3} = -V_n\ , 
\end{equation}
 which   gives a new reason why the NUT and Bolt spacetimes are somewhat complementary in their roles, like their (pun--fuelled\cite{Gibbons:1979xm}) namesakes, nuts and bolts.
 
Finally, we can define the internal energy for Taub--Bolt using $U=H-pV$, finding:
\begin{equation}
U_b=\frac{r_b^2+n^2}{2G r_b}\left( 1-\frac{3n^2}{\ell^2}\right)\ .
\end{equation}

\section{Closing Remarks}

We've found a very satisfying set of results for the physics of AdS--Taub--NUT and AdS--Taub--Bolt in the context of the extended thermodynamics where the cosmological constant defines a dynamical pressure. Our core idea is that the extended thermodynamics should not  be restricted to black hole spacetimes, but all gravitational spacetimes (with or without horizons), taking seriously the relationship between mass and enthalpy\cite{Kastor:2009wy}. This is particularly natural since there are ways of defining intrinsic thermodynamic properties for spacetimes that arise from  features that do not depend upon the presence of horizons, and the Taub--NUT and Taub--Bolt spacetimes are classic examples of this. 

Among the things we discovered were new examples of the thermodynamic volume being distinct from the naive geometric volume, as found for the rotating case in ref.\cite{Cvetic:2010jb}.  The presence of nut charge generates contributions to the thermodynamic volume\footnote{Continuing back to Lorenztian signature, we note that our Taub-NUT and Taub-Bolt spacetimes  share with the rotating case the property that they are stationary instead of static. It would be interesting to explore whether this is directly connected to the distinction between geometric and thermodynamic volumes. (We thank an anonymous referee for raising  this point.) However, the fact that their thermodynamic volumes can be formally represented as a volume integral, as shown in subsection~\ref{eq:naive-volume}, shows that they have a lot in common with the static case.}.  For the Taub--NUT case, we even had the extreme case that the naive geometric volume is  zero, while the thermodynamic volume is in fact non--zero and  negative. In our dynamical setting, we interpreted\footnote{Ref.\cite{Xu:2013zea}  reported a negative sign for a thermodynamic volume in the  context of black holes in Gauss--Bonnet gravity, but gave no interpretation. Since an  enthalpy was defined there for an analogous extended thermodynamics based on the effective cosmological constant set by the Gauss--Bonnet parameter, our interpretation may well apply there too. We thank Wei Xu for informing us of this work.} the negative sign as simply the result of the environment (the rest of the universe) doing work on the system to create the solution, which corresponds to the universe having had to increase its volume. Mechanical work is entirely natural in this extended setting, as discussed and exploited in ref.\cite{Johnson:2014yja}.

The picture we emphasise in all of this is that {\it any} gravity solution should be able to be thought of as a result of a thermodynamic process, and its thermodynamic properties interpreted in that light. One of the key points that emerges from this point of view is that certain puzzling thermodynamic features of the spacetimes are potentially explained in this setting:  Thermodynamic volumes can change dynamically (in both directions), and heat flows  in and out of such volumes also now have natural meaning. This gives  a dynamical origin for negative entropy of a spacetime solution, for example. A core idea was to take seriously the relationship between mass and enthalpy in this dynamical setting  for spacetimes beyond just black holes. It is to be expected that further work along these lines in other spacetime examples will yield similarly illuminating   physics.

\bigskip

\section*{Acknowledgements}
  CVJ would like to thank the  US Department of Energy for support under grant DE-SC0011687,  the Aspen Center for Physics  for hospitality (under NSF Grant \#1066293) during the preparation of a later version of this manuscript, and Amelia for her support and patience.


\providecommand{\href}[2]{#2}\begingroup\raggedright\endgroup

\end{document}